\documentclass[11pt,a4paper]{article}
\usepackage{latexsym}

\def\hybrid{\topmargin -20pt    \oddsidemargin 0pt
        \headheight 0pt \headsep 0pt
        \textwidth 6.25in       
        \textheight 9.5in       
        \marginparwidth .875in
        \parskip 5pt plus 1pt   \jot = 1.5ex}

 \hybrid

\catcode`\@=11

\def\marginnote#1{}
\newcount\hour
\newcount\minute
\newtoks\amorpm
\hour=\time\divide\hour by60
\minute=\time{\multiply\hour by60 \global\advance\minute by-\hour}
\edef\standardtime{{\ifnum\hour<12 \global\amorpm={am}%
        \else\global\amorpm={pm}\advance\hour by-12 \fi
        \ifnum\hour=0 \hour=12 \fi
        \number\hour:\ifnum\minute<10 0\fi\number\minute\the\amorpm}}
\edef\militarytime{\number\hour:\ifnum\minute<10 0\fi\number\minute}

\def\draftlabel#1{{\@bsphack\if@filesw {\let\thepage\relax
   \xdef\@gtempa{\write\@auxout{\string
      \newlabel{#1}{{\@currentlabel}{\thepage}}}}}\@gtempa
   \if@nobreak \ifvmode\nobreak\fi\fi\fi\@esphack}
        \gdef\@eqnlabel{#1}}
\def\@eqnlabel{}
\def\@vacuum{}
\def\draftmarginnote#1{\marginpar{\raggedright\scriptsize\tt#1}}

\def\draft{\oddsidemargin -.5truein
        \def\@oddfoot{\sl preliminary draft \hfil
        \rm\thepage\hfil\sl\today\quad\militarytime}
        \let\@evenfoot\@oddfoot \overfullrule 3pt
        \let\label=\draftlabel
        \let\marginnote=\draftmarginnote
   \def\@eqnnum{(\theequation)\rlap{\kern\marginparsep\tt\@eqnlabel}%
\global\let\@eqnlabel\@vacuum}  }

\def\preprint{\twocolumn\sloppy\flushbottom\parindent 2em
        \leftmargini 2em\leftmarginv .5em\leftmarginvi .5em
        \oddsidemargin -.5in    \evensidemargin -.5in
        \columnsep .4in \footheight 0pt
        \textwidth 10.in        \topmargin  -.4in
        \headheight 12pt \topskip .4in
        \textheight 6.9in \footskip 0pt
        \def\@oddhead{\thepage\hfil\addtocounter{page}{1}\thepage}
        \let\@evenhead\@oddhead \def\@oddfoot{} \def\@evenfoot{} }

\def\numberbysection{\@addtoreset{equation}{section}
        \def\theequation{\thesection.\arabic{equation}}}

\def\underline#1{\relax\ifmmode\@@underline#1\else
        $\@@underline{\hbox{#1}}$\relax\fi}

\def\titlepage{\@restonecolfalse\if@twocolumn\@restonecoltrue\onecolumn
     \else \newpage \fi \thispagestyle{empty}\c@page\z@
        \def\thefootnote{\fnsymbol{footnote}} }

\def\endtitlepage{\if@restonecol\twocolumn \else \newpage \fi
        \def\thefootnote{\arabic{footnote}}
        \setcounter{footnote}{0}}  

\catcode`@=12
\relax

\makeatletter

\def\publist{\@ifnextchar[{\@publist}{\@@publist}}
\def\@publist[#1]{\list
        {[\arabic{pubctr}]\hfill}{\settowidth\labelwidth{[999]}
        \leftmargin\labelwidth
        \advance\leftmargin\labelsep
        \@nmbrlisttrue\def\@listctr{pubctr}
        \setcounter{pubctr}{#1}\addtocounter{pubctr}{-1}}}
\def\@@publist{\list
        {[\arabic{pubctr}]\hfill}{\settowidth\labelwidth{[999]}
        \leftmargin\labelwidth
        \advance\leftmargin\labelsep
        \@nmbrlisttrue\def\@listctr{pubctr}}}
 \relax
\makeatother
\newskip\humongous \humongous=0pt plus 1000pt minus 1000pt

\newif\ifdtup

\relax

\def\d{\partial}

\def\sqr#1#2{{\vcenter{\vbox{\hrule height.#2pt\hbox{\vrule width.#2pt
height#1pt \kern#1pt \vrule width.#2pt}\hrule height.#2pt}}}}

\def\=d{\,{\buildrel\rm def\over =}\,}

\def\F{{\cal F}}

\def\i3p{\p32\int d^3p}

\def\As{A\hbox to 1pt{\hss /}}
\def\np4{\int d^4p_1\cdots d^4p_{n-1}\, }

\def\nx4{\int d^4x_1\ldots d^4x_n\, }

\def\kon#1#2{\vbox{\halign{##&&##\cr
\lower4pt\hbox{$\scriptscriptstyle\vert$}\hrulefill &
\hrulefill\lower4pt\hbox{$\scriptscriptstyle\vert$}\cr $#1$&
$#2$\cr}}}

\def\konv#1#2#3{\hbox{\vrule height12pt depth-1pt}
\vbox{\hrule height12pt width#1cm depth-11.6pt}
\hbox{\vrule height6.5pt depth-0.5pt}
\vbox{\hrule height11pt width#2cm depth-10.6pt\kern5pt
      \hrule height6.5pt width#2cm depth-6.1pt}
\hbox{\vrule height12pt depth-1pt}
\vbox{\hrule height6.5pt width#3cm depth-6.1pt}
\hbox{\vrule height6.5pt depth-0.5pt}}
\def\konu#1#2#3{\hbox{\vrule height12pt depth-1pt}
\vbox{\hrule height1pt width#1cm depth-0.6pt}
\hbox{\vrule height12pt depth-6.5pt}
\vbox{\hrule height6pt width#2cm depth-5.6pt\kern5pt
      \hrule height1pt width#2cm depth-0.6pt}
\hbox{\vrule height12pt depth-6.5pt}
\vbox{\hrule height1pt width#3cm depth-0.6pt}
\hbox{\vrule height12pt depth-1pt}}

\def\konw#1#2#3{\hbox{\vrule height12pt depth-1pt}
\vbox{\hrule height12pt width#1cm depth-11.6pt}
\hbox{\vrule height6.5pt depth-0.5pt}
\vbox{\hrule height12pt width#2cm depth-11.6pt \kern5pt
      \hrule height6.5pt width#2cm depth-6.1pt}
\hbox{\vrule height6.5pt depth-0.5pt}
\vbox{\hrule height12pt width#3cm depth-11.6pt}
\hbox{\vrule height12pt depth-1pt}}

\def\i{{\rm int}}

\def\e{{\rm ext}}

\def\r{{\rm ret}}
\def\a{{\rm av}}

\def\m3{{\mu_1\mu_2\mu_3}}

\def\co{{\rm Com}}

\def\p{{(+)}}

\def\be{\begin{equation}}       \def\eq{\begin{equation}}
\def\ee{\end{equation}}         \def\eqe{\end{equation}}

\def\bea{\begin{eqnarray}}      \def\eqa{\begin{eqnarray}}
\def\ena{\end{eqnarray}}        \def\eea{\end{eqnarray}}
                                \def\eqae{\end{eqnarray}}

\def\ba{\begin{array}}
\def\ea{\end{array}}
\def\unit{1 \hskip-.3em \raise2pt\hbox{$ \scriptstyle |$ } }

\def\a{\alpha}
\def\b{\beta}

\def\d{\delta}
\def\e{\epsilon}           

\def\h{\eta}
\def\i{\iota}

\def\l{\lambda}
\def\m{\mu}
\def\n{\nu}
\def\o{\omega}  
\def\p{\pi}                
\def\r{\rho}                                     
\def\s{\sigma}                                   
\def\t{\tau}

\def\F{\Phi}

\def\L{\Lambda}

\def\cd{{\cal D}}

\def\cf{{\cal F}}
\def\cg{{\cal G}}

\def\cl{{\cal L}}

\def\co{{\cal O}}

\def\half{{1 \over 2}}

\def\bop#1{\setbox0=\hbox{$#1M$}\mkern1.5mu
        \vbox{\hrule height0pt depth.04\ht0
        \hbox{\vrule width.04\ht0 height.9\ht0 \kern.9\ht0
        \vrule width.04\ht0}\hrule height.04\ht0}\mkern1.5mu}
\def\Box{{\mathpalette\bop{}}}                        
\def\pa{\partial}                              

\def\>{\rangle} 

\def\<{\langle} 
\def\Dsl{D \hskip-.6em \raise1pt\hbox{$ / $ } }

\def\sl#1{\rlap{\hbox{$\mskip 1 mu /$}}#1}
\def\leftrightarrowfill{$\mathsurround=0pt \mathord\leftarrow \mkern-6mu
       \cleaders\hbox{$\mkern-2mu \mathord- \mkern-2mu$}\hfill
       \mkern-6mu \mathord\rightarrow$}
\def\dvec#1{\vbox{\ialign{##\crcr
       \leftrightarrowfill\crcr\noalign{\kern-1pt\nointerlineskip}
       $\hfil\displaystyle{#1}\hfil$\crcr}}}          
\def\hook#1{{\vrule height#1pt width0.4pt depth0pt}}
\def\leftrighthookfill#1{$\mathsurround=0pt \mathord\hook#1
       \hrulefill\mathord\hook#1$}
\def\underhook#1{\vtop{\ialign{##\crcr                 
       $\hfil\displaystyle{#1}\hfil$\crcr
       \noalign{\kern-1pt\nointerlineskip\vskip2pt}
       \leftrighthookfill5\crcr}}}
\def\smallunderhook#1{\vtop{\ialign{##\crcr      
       $\hfil\scriptstyle{#1}\hfil$\crcr
       \noalign{\kern-1pt\nointerlineskip\vskip2pt}
       \leftrighthookfill3\crcr}}}

\def\sfrac#1#2{{\vphantom1\smash{\lower.5ex\hbox{\small$#1$}}\over
       \vphantom1\smash{\raise.4ex\hbox{\small$#2$}}}} 
\def\bfrac#1#2{{\vphantom1\smash{\lower.5ex\hbox{$#1$}}\over
       \vphantom1\smash{\raise.3ex\hbox{$#2$}}}}      
\def\afrac#1#2{{\vphantom1\smash{\lower.5ex\hbox{$#1$}}\over#2}}  
\def\on#1#2{{\buildrel{\mkern2.5mu#1\mkern-2.5mu}\over{#2}}}
\def\ddt#1{\on{\hbox{\LARGE .\kern-2pt.}}#1}             
\def\tdt#1{\on{\hbox{\LARGE .\kern-2pt.\kern-2pt.}}#1}   

\def\boxes#1{
       \newcount\num
       \num=1
       \newdimen\downsy
       \downsy=-1.5ex
       \mskip-2.8mu
       \bo
       \loop
       \ifnum\num<#1
       \llap{\raise\num\downsy\hbox{$\bo$}}
       \advance\num by1
       \repeat}
\def\boxup#1#2{\newcount\numup
       \numup=#1
       \advance\numup by-1
       \newdimen\upsy
       \upsy=.75ex
       \mskip2.8mu
       \raise\numup\upsy\hbox{$#2$}}

\newskip\humongous \humongous=0pt plus 1000pt minus 1000pt

\newif\ifdtup

\def\PRD{Phys. Rev. D}

\def\NPB#1#2#3{{\it Nucl. Phys.} {\bf B#1} (#2) #3}

\def\PRD#1#2#3{{\it Phys. Rev.} {\bf D#1} (#2) #3}

\def\PLB#1#2#3{{\it Phys. Lett.} {\bf #1B} (#2) #3}

\def\IJMPA#1#2#3{{\it Int. J. Mod. Phys.} {\bf A#1} (#2) #3}

\def\MPLA#1#2#3{{\it Mod. Phys. Lett.} {\bf A#1} (#2) #3}

\def\1ov4{{1\over 4}}

\def\pa{\partial}

\def\ddt{\dot{\t}}

\def\ta{\tilde{\a}}
\def\tb{\tilde{\b}}

\def\pa{\partial}

\def\half{{1 \over 2}}

\begin{document}

\begin{titlepage}
\begin{center}
\hfill KUL-TF-98/47\\
\hfill VUB/TENA/98/11\\ [3mm]
\hfill {\tt hep-th/9812207}\\

\vskip 5mm

{\bf Dilaton transformation under abelian and
non-abelian T-duality in the path-integral approach}

\vskip .2in

{\bf J. De Jaegher$^a$, J. Raeymaekers$^a$, A. Sevrin$^b$ and
W.Troost$^a$}\footnote{\tt
Jeanne.DeJaegher@fys.kuleuven.ac.be, Joris.Raeymaekers@fys.kuleuven.ac.be,\\
asevrin@tena4.vub.ac.be, Walter.Troost@fys.kuleuven.ac.be}
\\
\vskip 1.2cm

$^a${\em Instituut voor theoretische fysica, Katholieke
Universiteit Leuven,}\\
{\em Celestijnenlaan 200D, B-3001 Leuven, Belgium}\\
$^b${\em Theoretische Natuurkunde, Vrije Universiteit Brussel,}\\
{\em Pleinlaan 2, B-1050 Brussels, Belgium}\\

\vskip .1in

\end{center}

\vskip .2in

\begin{center} {\bf ABSTRACT } \end{center}
\begin{quotation}\noindent
We present a convenient method for deriving the
transformation of the dilaton under T-duality in the
path-integral approach. Subtleties arising in
performing the integral over the gauge fields are
carefully analysed using Pauli-Villars regularization,
thereby clarifying existing ambiguities in the literature.
The formalism can not only be applied to the abelian case,
but, and this for the first time, to the non-abelian case as well.
Furthermore, by choosing a particular gauge, we directly obtain
the target-space covariant expression for the dual geometry
in the abelian case.
Finally it is shown that the conditions for gauging non-abelian isometries
are weaker than those generally found in the literature.

\end{quotation}

\vskip 1cm

\end{titlepage}
\vfill
\eject

\renewcommand{\theequation}{\thesection.\arabic{equation}}

\newpage

\setcounter{equation}{0}
\section{Introduction}

This paper is devoted to the clarification of several technical aspects of
T-duality in bosonic string models in the path-integral approach (for a
review, see \cite{review1}, \cite{review2}).

The paper consists
of three parts, in the first of which we elucidate the derivation of the
dilaton transformation under (abelian) T-duality. The setup is the usual one:
one
gauges an isometry of a bosonic sigma model and introduces a Lagrange
multiplier to constrain the corresponding field strength to vanish.
Integrating over two different sets of variables in the path integral
yields the original and the dual model, related by the
so-called Buscher rules. The transformation
of the metric and torsion potential comes from the classical
contribution of the integration over the gauge fields. The dilaton
transformation however is more subtle since it is a `quantum effect'
coming from a functional
determinant which should be carefully regularized. A first attempt to account
for the dilaton transformation from the path-integral point of view was made
by Buscher in \cite{Buscher}. However, repeating this calculation, one
finds that a term is missing from the result for the functional
determinant. This term cannot be absorbed in a shift of the dilaton or the
other background fields.
We resolve this apparent catastrophe  by carefully retracing
the steps which lead from the gauged model back to the original one: one
encounters then another functional determinant which in \cite{Buscher} was
taken to be field independent, while in \cite{Tseytlin1,Tseytlin3} it was
realized
that this determinant should be regularized in a field dependent way so as
to cancel the unwanted term. The presentation  in \cite{Tseytlin1,Tseytlin3}
was
however restricted to a special class of sigma models, characterized by a
static metric and zero torsion, while here it is obtained for a generic
background.

For this purpose, we develop an unambiguous way to deal
with the regularization of these functional determinants through the use
of
Pauli-Villars (PV) regularization:
divergent Feynman diagrams are regularized by introducing extra massive fields
called PV fields. It has the advantage that these fields enter as extra terms
in the Lagrangian and are fixed once and for all. As a consequence, the
dependence of the path integral measure on parameters such as the conformal
factor
of the 2d metric, can be read off directly
from the mass terms of the PV fields \cite{Troost2}. Specifically, we use
 PV fields to regularize the gauged sigma-model. We then proceed to compute
both original and dual partition functions, and we obtain the
usual transformation of the dilaton field. We argue that the dilaton
transformation is a regularization-independent effect not affected
by ambiguities in the definition of the path integral measure for the
gauged model.

In the second part of the paper we tackle the covariant derivation.
We introduce a way to fix the gauge in a universal way and
obtain the dual model expressed in arbitrary coordinates. These covariant
T-duality rules facilitate the discussion of global issues like scaling and
singularities
of the dual model. Though the
coordinate-independent form of the Buscher rules was previously derived
in the Hamiltonian formulation \cite{cantransf}, in which T-duality is seen as
a canonical
transformation, quantum effects are more easily calculable in our approach.

One can also consider target spaces which have several non-commuting
isometries.
The dualization procedure can be generalized to these
backgrounds and is called non-abelian T-duality \cite{quevedo},
\cite{alvarez}.
The dilaton transformation
for this case has not yet been derived from first principles in the
literature.
In the third part of the paper we present such a derivation following
the same method as  in the abelian case. The starting point for this
derivation is the action with gauged non-commuting isometries.
The gauging
of non-abelian isometries presented in \cite{Hull}
has the disadvantage of not being applicable to the abelian case. We have
found that
it is possible to relax the conditions on target space given in \cite{Hull}
such that the
gauging of non-commuting isometries includes this case.

\setcounter{equation}{0}
\section{Abelian T-duality}

In this section we will demonstrate in detail how to obtain, through a
regulazised
computation, the T-dual partition
functions for closed bosonic strings, including the quantum corrections.

A standard method to perform T-duality is the following:
one gauges an isometry of a bosonic sigma-model and introduces
a Lagrange multiplier. Integrating over the Lagrange multiplier yields back
the original theory,
while integrating over the gauge fields brings up the T-dual theory.
Both describe equivalent string theories, where the strings
move in different background fields which are related by the
Buscher rules. These include a quantum effect, a shift of the dilaton field.
The necessity for this shift is deduced from the observation that, when the
original background fields solve the first order $\beta$-functions, the
classical dual background fields may not (for a recent discussion,
 see e.g. \cite{Balog}). In this section,
 we provide a fully regularized synthetic computation of this effect.

Let us start from the usual action for a closed bosonic
string in  a non-trivial background\footnote{We denote local world-sheet
coordinates
by $\s^a$ and local target space coordinates by $X^\m = (X^0, X^M)$.}
$(G_{\mu\nu}, B_{\mu\nu}, \Phi)$ with vanishing 1-loop $\beta$-functions.
Suppose the action is
invariant under the global tranformation $$\delta X^0 = \epsilon\,.$$ and that
none of the background
fields depends upon this coordinate. We can gauge this isometry by introducing
a gauge field
$A_a$ and a Lagrange multiplier $\chi$ which constrains the corresponding
field strength to vanish.
  The starting point for obtaining two dual actions is the following partition
function:
\begin{equation}
Z[G_{\mu\nu},B_{\mu\nu},\Phi,X^M]=\int \frac{[dA_a][dX^0][d\chi]}
{\rm gauge \  volume}
e^{-S_{gauged}} 
.\label{partfun1}
\end{equation}
The $X^M$ are treated as classical background fields throughout.

The gauged action reads
 \begin{eqnarray}
 S_{gauged} &=&\frac{1}{4\pi\alpha'}\int
d^2\sigma\sqrt{g}\bigg\{ (g^{ab}G_{MN}+
i\frac{\epsilon^{ab}}{\sqrt{g}} B_{MN})
\partial_a X^M\partial_b X^N \nonumber\\
&& + G_{00}g^{ab}(\partial_a X^0 + A_a)(\partial_b X^0 + A_b)\nonumber \\
&& + 2(g^{ab}G_{0M}+i\frac{\epsilon^{ab}}{\sqrt{g}}
B_{0M})(\partial_a X^0 +A_a)\partial_b X^M\nonumber \\
&&   +2i \frac{ \epsilon^{ab}}{\sqrt{g}} \chi \partial_a A_b + \alpha' R \Phi
\bigg\}.\label{action1}
\end{eqnarray}

The original stringmodel is obtained by integrating over the Lagrange
multiplier $\chi$, so that the gauge field becomes pure gauge, and gauge
fixing it to 0. This leaves the $X^0$ integral. An alternative way, which
makes clearer the procedure to obtain the dual theory, and opens the way to
treat
functional determinants less cavalierly, is to parametrize the
gauge field  $A_a$ in terms of two scalars $\ta, \tb$  (Hodge decomposition)
\be
A_a=\partial_a \ta +
\frac{\epsilon_a^{\;c}}{\sqrt{g}}\partial_c\tb.
\label{Aparam'}
\ee
The gauged isometry of the action translates into the fact that it depends on
the combination $X_0+\ta$, so that gauge fixing is performed simply by fixing
either $X_0$ or
$\ta$ to some convenient value. The action for the Lagrange multiplier
$\chi$ and the remaining gauge field degree of freedom $\tb$ contains a {\em
background independent} d'Alembertian, so that integrating out these fields
gives, at
first sight, just a constant --- but we will see that this naive expectation
is not valid
if we adopt the most straightforward regularization that respects
2-dimensional diffeomorphism invariance.

The dual model is obtained most simply by shifting the $A_a$-field after
completing the square, while fixing the gauge on  $X_0$. We find it most
convenient
to choose the coordinate equal to the Lagrange multiplier
\begin{equation}
 X^0 = \chi ,\label{gauge}
\end{equation}
instead of putting it to zero (as is often done):
this shows more explicitly how the Lagrange
multiplier $\chi$ becomes a coordinate of the dual target space%
\footnote{In the next section, the coordinate free formulation of
this gauge choice will be shown to lead to the covariant form of the dual
geometry.}.
The resulting action is
\begin{eqnarray}
 S_{gauged} &=&\frac{1}{4\pi\alpha'}\int d^2\sigma \sqrt{g}
\bigg\{ G_{00} g^{ab} A_a A_b\nonumber\\
&& + (g^{ab}\tilde{G}_{MN}+i{\epsilon^{ab} \over \sqrt{g}}\tilde{B}_{MN})
\partial_a X^M\partial_b X^N +\alpha' R \Phi\nonumber\\
   && +2 (g^{ab}\tilde{G}_{0M}+i{\epsilon^{ab} \over \sqrt{g}}\tilde{B}_{0M})
\partial_a\chi\partial_b X^M + g^{ab}\tilde{G}_{00}\partial_a\chi
 \partial_b \chi\bigg\},\label{action3}
\end{eqnarray}
which, apart from the quadratic gauge field term, is recognised to be a
bosonic sigma-model
(with $\chi$ interpreted as the zeroth coordinate) in the
background ($\tilde{G}_{\mu\nu}$, $\tilde{B}_{\mu\nu}$, $\Phi$):
\begin{eqnarray}
\tilde G_{00}&=&\frac{1}{G_{00}},\nonumber\\
\tilde G_{0M}&=&\frac{B_{0M}}{G_{00}},\nonumber\\
\tilde G_{MN}&=&G_{MN}-\frac{G_{0M}G_{0N}-B_{0M}B_{0N}}{G_{00}},\nonumber\\
\tilde B_{0M}&=&\frac{G_{0M}}{G_{00}},\nonumber\\
\tilde B_{MN}&=&B_{MN}-\frac{G_{0M}B_{0N}-B_{0M}G_{0N}}{G_{00}}.
\label{noncovdualfield}
\end{eqnarray}
Thus one obtaines the classical Buscher rules for the relation
between the metrics. To compute the quantum corrections, we introduce a
regularization.
\subsection{Regularization}
Since all functional integrals we want to perform are gaussian, regularization
 is rather
straightforward. We adopt a method dating back to Pauli and Vilars.
It can be viewed as adding extra fields (PV-fields) to the theory that have
very large
masses. In loops, they cancel the divergences of the original fields.
Physical results are obtained by letting the masses tend to infinity, possibly
after adding further terms to the action to make the limit finite.
Consistency on comparison of the original and the dual model will be achieved
only if we specify the {\em same} PV-action to provide the regularization in
both computations.

The folowing recipe constructs a PV action that is guaranteed to regularize
all one-loop diagrams (with external $\phi$-lines) \cite{Troost2}:
\begin{itemize}
\setlength{\itemsep}{-5pt}
\item take the second derivative $\frac{\partial^2 S(\phi)}{\partial
\phi(x)\partial\phi(x')}$
of the action $S(\phi)$ for the ordinary physical fields $\phi$;
\item sandwich this matrix with PV fields $\Phi$, one for every $\phi$;
thus,  PV fields have the same kinetic term as the field
they regularize.
\item choose a mass term $M^2 \Phi T \Phi$, with $T$ a (non-degenerate) matrix
that may depend on the fields
$\phi$.
\item add these constructs to the action; include a minus sign for diagrams
with a closed PV-loop.
\footnote{If one wishes, the minus sign prescription may be
dispensed with at the expense of introducing three fields, two fermionic and
one bosonic,
in stead of the one "bosonic" PV-field.
Since for the present purpose this would not change anything,
we stick to the simpler representation.}.
\item if necessary for the regularization of momentum integrals, add {\em
several} sets of
PV-fields with identical
actions but values $M^2_j$ for the masses, weighing the loops with
factors $c_j$: effectively, integration over PV fields is defined through
\be
\int [d\Phi_j] e^{- \int d^2 \sigma \Phi_j A \Phi_j} = (det A)^{\frac{c_j}{2}}
.
\label{PVweights}
\ee
One imposes the regularization conditions
\begin{eqnarray}
\sum_j c_j &=& 1,\nonumber\\
\sum_j c_j M_j^2 &=& 0. \label{PVcond}
\end{eqnarray}
\end{itemize}

Implementing this for the action (\ref{action1}) with the change
of variables (\ref{Aparam'}), where three PV fields are introduced
corresponding to the fields $\{X_0+\ta,\tb,\chi\}$, leads
to an awkward non-diagonal action. A considerable simplification occurs if we
decompose the gauge field $A_a$ in a slightly different  way, viz.
\be
A_a=\partial_a \alpha +
\frac{\epsilon_a^{\;c}}{\sqrt{g} G_{00}}\partial_c\beta.
\label{Aparam}
\ee
That such a decomposition is always possible locally can be argued
as follows: one first defines a symmetric, positive bilinear inner product
on the space of $n$-forms by
\be
( \o, \h ) = \int G_{00} \,\o \wedge \star \,\h.
\label{inprod}
\ee
This defines the hermitean conjugate $d^\dagger$ of the operator $d$.
A generalised
Hodge decomposition theorem then states that locally every form $\o$ can be
decomposed into an exact and a coexact part. This decomposition is
orthogonal with respect to the inner product (\ref{inprod}). Applying
this to the one-form $A$ we get the decomposition (\ref{Aparam}).
The parametrization (\ref{Aparam}) has the advantage that the
$\a, \b$ cross-terms vanish so that the kinetic energy operator for
$\a$ and $\b$ is diagonal.

We now introduce the (sets of) Pauli-Villars fields ${Y}^i_0$,
$Y^i_1$, $Y^i_2$ to regularize integrations over $\chi$, $\alpha$ and $\beta$
respectively. These are the only integrations that need to be performed
for comparison of the dual versions. We choose the mass terms to respect
worldsheet reparametrization invariance, which however entails breaking
two dimensional conformal invariance.

{}From (\ref{action3}) and (\ref{Aparam}) we find
\begin{eqnarray}
S_{PV}[ Y^i_0]&=&\frac{1}{4\pi\alpha'}\sum_i\int d^2\sigma \sqrt{g}
\, \frac{1}{G_{00}}(g^{ab} \partial_a Y^i_0 \partial_b Y^i_0 + M_i^2
(Y^i_0)^2), \nonumber\\
S_{PV}[Y^i_1]&=& \frac{1}{4\pi\alpha'}\sum_i\int d^2\sigma \sqrt{g}
\, G_{00}(g^{ab} \partial_a Y^i_1 \partial_b Y^i_1 + M_i^2
(Y^i_1)^2), \nonumber\\
S_{PV}[Y^i_2]&=&\frac{1}{4\pi\alpha'}\sum_i\int d^2\sigma \sqrt{g}
\, \frac{1}{G_{00}}(g^{ab} \partial_a Y^i_2 \partial_b Y^i_2 + M_i^2
(Y^i_2)^2) .
\label{PVterms}
\end{eqnarray}
The path-integral measure is now explicitly defined\footnote{One might
consider including a (formal) Jacobian factor  for the change of
variables from $A_a$ to $\a, \b$. This would amount to no  more than
a change of normalization of the partition
function for the gauged sigma model and  drops out of the comparison
of the original to the dual partition function. We simply adopt
(\ref{measure}) as our definition.}
by
\begin{eqnarray}
\frac{[dX_0][dA_a][d\chi]}{\rm gauge \  volume}
&\equiv&
\prod_\sigma
 \Big( d(X_0(\sigma)+\alpha (\sigma))  \prod_j d(Y_1^j(\sigma))\nonumber\\
&&\quad\quad d(\beta(\sigma)) \prod_k d(Y_2^k(\sigma))\nonumber\\
&&\quad\quad d(\chi (\sigma))\prod_i d(\ Y^i_0 (\sigma))
\Big).
\label{measure}
\end{eqnarray}
The path integral measures in (\ref{measure}) are (line by line) invariant
under
world-sheet reparametrizations and Weyl rescalings: because of the
"statistics" of the PV fields (compare (\ref{PVweights}) ) the
Jacobian for transformation of a field  cancels with the
transformation of the corresponding PV-field (see for example
\cite{Troost2}). Hence, with this choice of mass terms, the
partition function is invariant under reparametrizations of the
world-sheet, and a possible breaking of Weyl invariance comes
exclusively from the mass terms.

The world sheet reparametrization invariance permits us to
(locally) choose a conformal gauge: $$g_{ab}=\rho\delta_{ab}\,.$$
Loop effects can be accessed via the dependence of the functional integrals on
this conformal factor.

\subsection{The quantum contributions}
\subsubsection{In the original formulation of the model}
The original formulation of the sigma model is obtained by integrating over
the gauge field
and the Lagrange multiplier $\chi$, leaving the $X^0$ integral. Starting from
the gauged
action (\ref{partfun1}), we set $\alpha = 0$.
This gauge choice implies that the $X^0$ integral is regularized by
the set of Pauli-Villars fields $Y^i_1$.
The $\chi$ and $\beta$ integrations are carried out,
together with the integrals over ${Y}^i_0$ and $Y^i_2$ that regularize them.
The resulting path integral contains first of all  the bosonic
string  sigma-model action in the original  background
($G_{\mu\nu}$, $B_{\mu\nu}$, $\Phi$), with isometry coordinate $X^0$,
 the remaining part is the regularized integral over $\beta$ and $\chi$:
\be
Z= \int [dX^0]\prod_{i}[dY^i_1]\;\exp\Big(-S[G_{\mu\nu},B_{\mu\nu},
\Phi,X^\mu]- S_{PV}[Y_1^i]  \Big) \ \  e^{-W}
\ee
with
\bea
e^{-W}&=& \int  [d\beta][d\chi]\prod_{j}[dY^j_0][dY^j_2]\; \exp\bigg(
- S_{PV}[Y_0^j] - S_{PV}[Y_2^j] \nonumber \\
       &\quad & \quad -\frac{1}{4\pi\alpha'}\int d^2\sigma
\bigg\{ {1 \over G_{00}} \delta^{ab}( \partial_a \beta\partial_b\beta
+ 2 i  \partial_a \chi \partial_b \b)
\nonumber \\
       &\quad &\quad +
2(\delta^{ab}G_{0M}+i\varepsilon^{ab}B_{0M})\partial_b X^M
{\varepsilon_a^{\;c} \over G_{00}}
\partial_c\beta \bigg\}\bigg)\,.
\end{eqnarray}
The kinetic energy  for $\b$ and $\chi$ can be diagonalised
by completing the square in $\b$ and defining a shifted variable
$\b' = \b + i \chi$:
\bea
e^{-W}&=& \int  [d\beta'][d\chi]\prod_{j}[dY^j_0][dY^j_2]\;\exp\bigg(
- S_{PV}[Y_0^j] - S_{PV}[Y_2^j] \nonumber \\
&\quad & \quad -\frac{1}{4\pi\alpha'}\int d^2\sigma
\bigg\{ {1 \over G_{00}} \delta^{ab}( \partial_a \beta'\partial_b\beta'
+ \partial_a \chi \partial_b \chi)\nonumber\\
&\ &- 2 (\b' - i \chi) \varepsilon_a^{\ c}\partial_c
[(\delta^{ab}{G_{0M} \over G_{00}}+i\varepsilon^{ab}{B_{0M}\over
G_{00}})\partial_b X^M]
 \bigg\} \bigg)\,.
\eea
Integrating over $\b'$, $\chi$ and their corresponding PV fields,  we obtain
\begin{equation}
W[\rho,G_{00}]=-\sum_i \frac{c_i}{2}{\rm Tr}\ln\left(\frac{\co + \rho
M_i^2}{{\co}}\right)\,,
\end{equation}
with
\be
\co = \left( \begin{array}{cc} \delta^{ab}(-\partial_a\partial_b-\Lambda_a
\partial_b) & 0\\ 0&  \delta^{ab}(-\partial_a\partial_b-\Lambda_a\partial_b)
\end{array}\right)\,,
\ee
and
\begin{equation} \Lambda_a=\partial_a \ln G_{00} \,. \end{equation}
The variation of $W$ with respect to the conformal factor is calculated in
appendix \ref{app1} by using the heat-kernel expansion \cite{gilkey}.
The result can be integrated with respect to $\r$ to give
(see \ref{origcontrib})
\be
W =  {1 \over 4 \p} \int d^2 \s \sqrt{g} [ {1 \over 12}
R \Box^{-1} R + {1 \over 4}\Box^{-1} R \; \L^2 + \half R \ln G_{00}] +
{\rm Weyl \ inv.\ terms}.
\label{W}
\ee
The first term can be seen as an extra contribution to the Liouville action
coming from the gauge field and the Lagrange multiplier.
Note the dependence of (\ref{W}) on $G_{00}$, that, with this regularization,
comes in
through the functional integrals leading to the {\em original} sigma model.
This dependence may be missed if one does these integrals naively, and in
particular the term quadratic in $\Lambda$ is absent in \cite{Buscher}.
The term with $\ln G_{00}$ will give rise to the dilaton shift; the other
terms
can not be absorbed in the dilaton, but will eventually drop out.

The computation via the conformal anomaly gives no control over possible Weyl
invariant terms \cite{Tseytlin3}. However, these are restricted to  be of the
form  $2 c_1 \int
d^2 \s \sqrt{g}
 g^{ab}\Lambda_a\Lambda_b$, with an undetermined coefficient $c_1$
from each operator $\delta^{ab}(-\partial_a\partial_b-\Lambda_a
\partial_b)$.

Summarizing,
 \begin{eqnarray}
Z &=&\exp\left( -{1 \over 4 \p} \int d^2 \s \sqrt{g} [ {1 \over 12}
R \Box^{-1} R + {1 \over 4}\Box^{-1} R\; \L^2 + \half R \ln G_{00}
+ 2 c_1 \L^2]\right)
\nonumber \\[0,3cm]
       &\quad&\int [dX^0]\prod_{i}[dY^i_1]\; \exp\Big(-S[G_{\mu\nu},
B_{\mu\nu},\Phi,X^\mu]- S_{PV}[Y_1^i] \Big)\,.
\label{origpartfun2}
\end{eqnarray}

\subsubsection{In the dual formulation}
The dual model is obtained after integration over $\alpha$ and
$\beta$. The resulting partition function reads
\begin{equation}
Z =\int
[d\chi]\prod_{i}[dY^i_0]\exp\Big(-S[\tilde{G}_{\mu\nu},
\tilde{B}_{\mu\nu},\Phi] - S_{PV}[Y_0^i] \Big)\ \ e^{-\tilde W},\nonumber\\
\end{equation}
with
\begin{equation}
\tilde W[\rho,G_{00}]=-\sum_j \frac{c_j}{2}{\rm Tr}\ln\left(
\frac{\tilde \mathcal{O}+\rho M_j^2}{\tilde \mathcal{O}}
\right)\,,
\end{equation}
where
$$\tilde \mathcal{O}=\left( \begin{array}{cc}
\delta^{ab}(-\partial_a\partial_b-\Lambda_a
\partial_b) & 0\\ 0&  \delta^{ab}(-\partial_a\partial_b+\Lambda_a\partial_b)
\end{array}\right)\ .
$$
The variation  of $\tilde W$ with respect to  the conformal factor is
calculated in appendix \ref{app1}, again using the heat-kernel expansion.
 The result can be integrated with respect to $\r$ to give (\ref{dualcontrib})
\be
\tilde{W} =  {1 \over 4 \p} \int d^2 \s \sqrt{g} [ {1 \over 12}
R \Box^{-1} R + {1 \over 4}\Box^{-1} R \; \L^2] + {\rm Weyl \ inv.\ terms}.
\label{tildeW}
\ee
A comparison with Buscher's computation \cite{Buscher} done in the appendix,
shows that the $\L^2$ was not present in his results although it sould have
been (\ref{Buscherresult}).
Since the Weyl invariant term is symmetric under $\L
\rightarrow -\L$, its contribution is again $2 c_1 \int
d^2 \s \sqrt{g}
 g^{ab}\Lambda_a\Lambda_b$.
The resulting partition function at the dual side consists of the integrated
anomaly
 and the classical dual action:
\begin{eqnarray}
Z &=& \exp \Big(- {1 \over 4 \p} \int d^2 \s \sqrt{g} [ {1 \over 12}
R \Box^{-1} R + {1 \over 4}\Box^{-1} R \; \L^2 + 2 c_1 \L^2]\Big)\cdot
\nonumber\\[0,3cm]
&\quad&\int [d\chi]\prod_{i}[dY^i_0]\exp\Big(-S[\tilde{G}_{\mu\nu},
\tilde{B}_{\mu\nu},\Phi, \chi, X^M] - S_{PV}[Y_0^i]\Big).\label{dualpartfun}
\end{eqnarray}

\subsubsection{The dilaton shift}
Comparing the results in (\ref{origpartfun2}) and (\ref{dualpartfun}), the
Liouville action and the terms quadratic in $\Lambda$ are identical on both
sides. This is just as well, since they could not be absorbed
in any of the background fields, and would therefore constitute a breaking
of duality at the quantum level.
The difference that remains resides in a term that can be absorbed in a
shift of the dilaton field on either side:

\begin{equation}\tilde \Phi = \Phi - \frac{1}{2} \ln G_{00}
\;.\label{dilshift}\end{equation}

Summarizing, the identity
\begin{eqnarray}
&& \int [dX^0]\prod_{i}[dY^i_1]\,\exp (-S[G_{\mu\nu},B_{\mu\nu},\Phi,X^\m] -
S_{PV}[Y^i_1])\nonumber\\
&=&\int [d\chi]\prod_{i}[d Y^i_0]\;\exp (- S[\tilde
G_{\mu\nu},\tilde B_{\mu\nu},\tilde \Phi, \chi, X^M]- S_{PV}[{Y}^i_0])
\label{result}
\end{eqnarray}
shows the equivalence of the bosonic strings propagating
in dual backgrounds, both having a shift in the zeroth coordinate as global
isometry.

It is important to note that this
equivalence is valid {\em to all orders}, on condition that
the integration over the isometry coordinate is regularized as indicated by
the Pauli-Villars actions present in (\ref{result}), spelled out explicitly
in (\ref{PVterms}). If one prefers to perform the regularization differently,
one may have to compensate for this with an additional finite counterterm.
In the proof of this equivalence, we have kept the other string coordinates
$X^M$
fixed. If one also considers quantum fluctuations of these fields, they may
give further quantum contributions to (for example) beta functions, but, {\em
provided} one regularizes them in identical ways on both sides, this is
irrelevant for the T-duality rules  (\ref{dilshift}) and
(\ref{noncovdualfield})) as computed here.
However, if one adopts a regularization of the $X^M$ coordinates on the
original side that is
different from the dual side, counterterms have to
be introduced in the form of further corrections to the T-duality rules as in
\cite{Tseytlin1}, \cite{kaloper}.

\setcounter{equation}{0}
\section{Target-space covariant abelian T-duality rules}

We now proceed with the derivation of the Buscher
rules in a target space covariant way. After the construction of the action
where
one or several commuting global isometries are gauged, we concentrate
on the gauge fixing that is needed for dualizing the model at the quantum
level and present the
Buscher rules in a covariant form. While the classical rules were already
obtained
in \cite{cantransf} using canonical transformations, the quantum treatment was
not done
up till now as no suitable gauge fixing was available.

\subsection{The gauging}
Consider again the action for a bosonic string in non-zero (D)-dimensional
background fields
\begin{equation}
S=\frac{1}{4\pi\alpha'}\int
d^2\sigma\sqrt{g}\big\{(g^{ab}G_{\mu\nu}+i\frac{\epsilon^{ab}}
{\sqrt{g}}B_{\mu\nu})\partial_a
X^{\mu}\partial_b X^{\nu}+\alpha' R^{(2)}\Phi\big\}\,.
\label{sigma}
\end{equation}
Following \cite{Hull}, the transformation $\delta X^\mu=\varepsilon
k^\mu$ is a symmetry of $S$ if
\begin{itemize}
\item $k^\mu$ is a Killing vector: \begin{equation}\mathcal{L}_k G_{\mu\nu}=
k^\rho
G_{\mu\nu,\rho}+k^\rho_{\;,\mu}G_{\rho\nu}+k^\rho_{\;,\nu}G_{\mu\rho}=0.
\end{equation}
\item The Lie derivative of $B_{\mu\nu}$ is a total derivative\footnote{The
notation $i_k\omega$, with $k$ a vector
and $\omega$ a form stands for $i_k \omega = \frac{1}{n!}\sum_{s=1}^{n}
k^{\mu_s}
\omega_{\mu_1\cdots\mu_s\cdots\mu_n}
(-1)^{s-1} dx^{\mu_1}\wedge\cdots dx^{\mu_{s-1}}\wedge
dx^{\mu_{s+1}}\wedge\cdots dx^{\mu_n}$}:
 \begin{equation}\mathcal{L}_k B\equiv i_kdB+di_kB=d(v+i_kB),\end{equation}
 where $dv=i_k dB$
\item The dilaton is invariant:
\begin{equation}\mathcal{L}_k
\Phi=k^\mu\partial_\mu\Phi=0.\end{equation}
\end{itemize}

Promoting the isometry to a gauge symmetry, $\delta X^\mu=\varepsilon(\sigma)
k^\mu$, and applying
the Noether procedure results in an invariant action,
\begin{eqnarray}
S_{gauged}&=&\frac{1}{4\pi\alpha'}\int d^2\sigma
\bigg\{\sqrt{g}g^{ab}G_{\mu\nu}
(\partial_a X^\mu +A_ak^\mu)(\partial_b X^\nu +A_bk^\nu)\nonumber \\
&\qquad & +i\epsilon^{ab}B_{\mu\nu}\partial_a X^\mu \partial_b
X^\nu -2i\epsilon^{ab}
A_a v_\nu \partial_b X^\nu \nonumber \\
&\qquad &  +\alpha' R^{(2)}\Phi\bigg\},\label{action4}
\end{eqnarray}
where we introduced a gauge field $A_a$ transforming as
\begin{equation}\delta A_a= - \partial_a\varepsilon\,.\end{equation}
Invariance is easily checked once it is realized that $v_\mu \partial_aX^\mu$
can be chosen to be invariant.
Indeed, under a gauge transformation $v_\mu \partial_aX^\mu$ transforms as a
total derivative,
\begin{eqnarray}
\delta \left( v_\mu \partial_aX^\mu\right)=\partial_a\left( \varepsilon k^\mu
v_\mu \right).
\end{eqnarray}
However, it is clear from the definition of $v$ that it is only defined up to
an exact one-form.
In this way we have an additional shift symmetry,
\begin{eqnarray}
\delta_{shift}\left(v_\mu \partial_aX^\mu\right)=\partial_a h,
\end{eqnarray}
with $h$ arbitrary. For a given $\varepsilon$ we can make $v_\mu
\partial_aX^\mu$ invariant
by making a compensating shift transformation with parameter
$h=-\varepsilon k^\mu v_\mu $.

The starting point for the duality transformation is this gauged action
supplemented with
a Lagrange multiplier term
\begin{equation}
S_{lm} = \frac{i}{2\pi\alpha'}\int d^2\sigma  \epsilon^{ab}A_a
\partial_b\chi\,,
\end{equation} so that the partition function reads:
\begin{equation}
Z[G_{\mu\nu},B_{\mu\nu},\Phi,X^\mu,g^{ab}]=\int
\frac{[d A_a][d\chi][d\eta]}
{\rm gauge\; vol.} e^{-S_{gauged}- S_{lm}}\,.
\end{equation}
The role of the $\eta$ variable
will be explained in the next paragraph.
As in the rest of the paper, we always assume a trivial world sheet topology.
Using the methods
developed in \cite{erik} and  \cite{Lozano}, our analysis could be generalized
to non-trivial
topologies as well.

\subsection{The gauge fixing for the dual theory}

The key ingredient of our analysis is the choice of a local gauge, meaning
that the gauge is
fixed using fields transforming without derivatives on the transformation
parameter. As a consequence, the gauge fixing determinant
will be a trivial normalization factor of the path integral having no
influence
on the computation of the other non-trivial determinants.

The Killing vector $k^\mu$ generates the abelian isometry. One can always
define a 1-form $\omega$
that satisfies
\begin{equation} \omega_\mu k^\mu = 1 \,.\end{equation}
One can take $\omega$ to be  exact
and write it
as $d$ of a coordinate in the direction of the isometry \cite{dewitt}:
\begin{equation} \omega_\mu dX^\mu = d\eta \,.\end{equation}
Our gauge choice is then
\begin{equation} \eta = \chi \,.\end{equation}
such that the Lagrange multiplier becomes the new isometry coordinate of the
dual model. This condition can always
be reached taking the transformation parameter to be
\begin{equation} \epsilon = -\eta + \chi \,.\end{equation} Integration
over the Lagrange multiplier will then cancel the infinite volume of the gauge
group.
Note that this gauge choice is simply the covariant counterpart of the
``natural'' gauge choice of
our first derivation (\ref{gauge}).

\subsection{The covariant Buscher rules}

The calculation is now exactly the same as in adapted coordinates. The result
is a set
of covariant abelian T-duality rules:
\begin{eqnarray}
\tilde G_{\mu\nu}&=&G_{\mu\nu}-
\frac{k_\mu
k_\nu -(v_\mu-\omega_\mu)( v_\nu-\omega_\nu)}{k^2},\nonumber\\[0,3cm]
\tilde B_{\mu\nu}&=&B_{\mu\nu}
-\frac{k_\mu (v_\nu-\omega_\nu) -k_\nu
(v_\mu-\omega_\mu)}{k^2},\nonumber\\[0,3cm]
\tilde\Phi &=& \Phi-\frac{1}{2}\ln k^2\,.\label{covdualfield}
\end{eqnarray}

This discussion is trivially generalised
to backgrounds with several commuting isometries.

\setcounter{equation}{0}
\section{Dilaton transformation for non-abelian duality.}

In this section we present a systematic derivation of the dilaton
transformation for non-abelian duality \cite{quevedo}\cite{alvarez},
i.e. T-duality with respect
 to a non-commuting group of isometries of the background. The
derivation will proceed along the same lines as in the abelian case.

\subsection{Conditions for non-abelian symmetries}

We start with a bosonic string sigma model on a background
which we now suppose to have
 $n$ independent isometry vectors $k_i^\m, i=1 \ldots n$ that satisfy the
algebra
\be
[ k_i, k_j ] = f^k_{\ ij} k_k,
\ee
where the $f^k_{\ ij}$ are the  structure constants of the isometry group
\footnote{Anomaly considerations require the isometry group to
have traceless generators in the adjoint representation \cite{alvarez}.}.
The action  is invariant under the global transformations
$\d X^\mu = \e^i k^\m_i$ if the following conditions are satisfied
\cite{Hull}:
\begin{itemize}
\item  $k^\mu_i$ are Killing vectors:
\be
\cl_i G = 0 .
\ee
\item The Lie derivative of $B=\half B_{\mu\nu}dX^\mu\wedge dX^\nu$ is a
total derivative:
\be
 \cl_i B = d( v_i + i_i B) ,
\ee
or
\be
i_i dB = d v_i .
\label{Bcond}
\ee
\item The dilaton is invariant:
\be
\cl_i \F = 0  .
\ee
\end{itemize}
We used the notation $\cl_i \equiv \cl_{k_i}$ and $i_i  \equiv
i_{k_i} $ to denote the Lie derivative and index contraction respectively.
 As before, the one-forms $v_i$ are only defined up to a closed
form. By taking a Lie derivative of (\ref{Bcond}) one also finds that
\be
\cl_i v_j = f_{\ ij}^k v_k + w_{ij} , \label{vtran}
\ee
with  $w_{ij}$ a set of unspecified closed one-forms. We also define, for
later use, the set of functions $c_{ij}$ by
$$
c_{ij}\equiv i_j v_i.
$$

\subsection{Gauging and further conditions}
We promote the isometries to a gauge symmetry, $\d X^\m = \e^i (\s)
k_i^\m$.
It is important to realize that gauging non-abelian isometries
imposes further conditions on the background \cite{Hull}.
These conditions are usually arrived at as follows: the freedom to shift
the $v_i$ by a closed form is used to put the forms $w_{ij}$ to zero.
This is not always possible, the integrability conditions to achieve this are
derived in \cite{Hull}). The Noether
procedure then leads to one further condition: $c_{(ij)}=0$. However, in the
previous section we showed that for abelian isometries no conditions on
$w_{ij}$ or $c_{ij}$ were required to gauge the model. This suggests that
the gauging conditions usually adopted are too restrictive
and can be relaxed to
include the models with abelian isometries. In order to arrive at these more
general conditions we refrain from fixing the shift freedom on the
$v_i$ and include such shifts as compensating gauge transformations of the
$v_i$, just
as we did previously in the abelian models. Under a gauge transformation,
$v_{j\mu }\partial_aX^\mu $ transforms as
\be
\d \left(v_{j\mu }\partial_aX^\mu\right) =
\e^i\left( f^k_{\ ij} v_{k\mu } + w_{ij\mu } \right)\partial_aX^\mu
+ c_{ji} \partial_a \e ^i.  \label{var1}
\ee
As $v_j$ is only defined modulo an exact one-form, we get an additional shift
symmetry
\begin{eqnarray}
\delta_{shift} \left(v_{j\mu }\partial_aX^\mu\right)=\partial_a h_j,
\end{eqnarray}
with the functions $h_j$ completely arbitrary.
A Noether procedure as in \cite{Hull} determines
the functions $h_j$.
While doing this one finds an additional requirement: the closed one-forms
$w_{ij}$ should be exact as well,
\begin{eqnarray}
w_{ij}=d\lambda_{ij}.
\end{eqnarray}
In fact this is only a condition on the antisymmetric part of $w_{ij}$, as one
has
\begin{eqnarray}
w_{(ij)}=dc_{(ij)}\quad\mbox{ or }\quad \lambda_{(ij)}=c_{(ij)},
\end{eqnarray}
which can be seen by combining the variation in eq. (\ref{var1}) with the
alternative expression
\be
\d \left(v_{j\mu }\partial_aX^\mu\right) =
\partial_a(\e^ic_{ji})-3\e^ik_i^\nu k_j^\rho
\partial_{[\mu }B_{\nu\rho]}\partial_aX^\mu.
\ee
In this way we find for given $\e^i$ the following compensating shift
transformation,
\be
h_j =-\epsilon^i\lambda_{ij} =- \e^i c_{(ij)} - \e^i \l_{[ij]}.\label{trf0}
\ee
As a consequence the combined gauge and compensating shift transformation of
$v_{j\mu }\partial_aX^\mu$ becomes
\begin{eqnarray}
\delta \left(v_{j\mu }\partial_aX^\mu\right)=\e^i f^k_{\ ij} v_{k\mu }
\partial_aX^\mu - \partial_a \e ^i\left(\lambda_{[ij]}+c_{[ij]}\right),
\end{eqnarray}
and $c_{ij}$ transforms homogeneously,
\begin{eqnarray}
\delta c_{ij}=\e^k\left(f^l{}_{ki}c_{lj}+ f^l{}_{kj}c_{il}\right).
\end{eqnarray}
The gauged action becomes
\begin{eqnarray}
S_{gauged}&=&\frac{1}{4\pi\alpha'}\int d^2\sigma
\bigg\{\sqrt{g}g^{ab}G_{\mu\nu}
(\partial_a X^\mu -A^i_ak_i^\mu)(\partial_b X^\nu -A^i_bk_i^\nu)\nonumber \\
&\qquad & +i\epsilon^{ab}B_{\mu\nu}\partial_a X^\mu \partial_b X^\nu
+2i\epsilon^{ab}
A^i_a v_{i\nu} \partial_b X^\nu \nonumber \\
&\qquad & -i\epsilon^{ab}(c_{[ij]} + \l_{[ij]})A^i_a A^j_b
+ \alpha' R^{(2)}\Phi\bigg\},\label{gauged}
\end{eqnarray}
which is invariant under the gauge transformations
\bea
\d X^\m &=& \e^i  k_i^\m,\nonumber\\
\d A_a^i &=& \pa_a \e^i + f^i_{\ jk} A_a^j \e^k,
\eea
provided $\lambda_{[ij]}$ transforms homogeneously as well,
\begin{eqnarray}
\delta\lambda_{[ij]}=\epsilon^kf^l{}_{ki}\lambda_{[lj]}+
\epsilon^kf^l{}_{kj}\lambda_{[il]}. \label{trf1}
\end{eqnarray}
This implies further conditions on the background.
The transformation of $\lambda_{[ij]}$ can be decomposed as
\begin{eqnarray}
\delta \lambda_{[ij]}=\epsilon^k{\cal L}_k\lambda_{[ij]}
+\delta_{shift}\lambda_{[ij]} .
\label{trf2}
\end{eqnarray}
The compensating shift transformation is immediately obtained from eqs.
(\ref{vtran})
and (\ref{trf0}) which give
\begin{eqnarray}
\delta_{shift}\lambda_{ij}=-\epsilon^k\left({\cal
L}_i\lambda_{kj}-f^l{}_{ij}\lambda_{kl}
\right).\label{trf3}
\end{eqnarray}
Combining eqs. (\ref{trf1}), (\ref{trf2}) and (\ref{trf3}) yields the extra
condition
\begin{eqnarray}
\cl_k \l_{[ij]}+\cl_{[j} \l_{|k|i]}   = - f^l{}_{ij} \lambda_{kl}+
f^l{}_{ki}\lambda_{[lj]}-f^l{}_{kj}\lambda_{[li]}. \label{gaugecond}
\end{eqnarray}
This condition generalises the one found in \cite{Hull} and incorporates
the abelian models for which it is trivially satisfied. One can also check
that  it is invariant under the shifts $v_i \rightarrow v_i + d f_i$ for
arbitrary functions $f_i$. The conditions obtained in  \cite{Hull} form a
special
solution to our conditions. Indeed, in \cite{Hull}, one required that the
one-forms $v_i$
could be chosen in such a way that $w_{ij}=0$, implying that $c_{(ij)}$ is a
constant which
had to vanish. In this case $\lambda_{ij}$ can be choosen to be zero and eq.
(\ref{gaugecond})
is trivially satisfied.

We proceed by adding the  Lagrange multiplier term to eq. (\ref{gauged}),
\be
S_{lm} =  \frac{i}{4\pi\alpha'}\int d^2\sigma
  \epsilon^{ab}F^i_{ab} \chi_i,
\ee
where the $\chi_i$ transform as
$$
\d \chi_i = -f^j_{\ ik}\chi_j \e^k .
$$

Again the starting point for the duality transformation is
the partition function for the gauged model
\be
Z[G_{\mu\nu},B_{\mu\nu},\Phi,X^M]=\int [dA_a^i][dX^i][d\chi_i]
B[f_i] \det \cf
e^{-S_{gauged} - S_{lm}},\label{partfun3}
\ee
with $X^\m = (X^i, X^M)$ and the $X^i$ are a subset of coordinates
parametrizing the orbits
of the isometry group, $B$ is a suitably chosen functional of some
 gauge-fixing functions $f^i$ and $\det \cf$ is
the corresponding Fadeev-Popov determinant.

\subsection{The classical T-dual theory}

As in the abelian case, the dual model is obtained by integrating out the
gauge fields and making a gauge choice which gives the Lagrange multipliers
the meaning of functions on target space. The same procedure
will be followed here.

The gauged action can be written as follows:
\be
S_{gauged}+ S_{lm} = S + {1 \over 4 \p \a '} \int d^2 \s \sqrt{g} [A^i_a
f^{ab}_{ij}
A^j_b + h^a_i A^i_a],
\ee
with
\bea
f^{ab}_{ij} &=& g^{ab} \cg _{ij} + {i \over \sqrt{g}} \e^{ab} \cd _{ij},\nonumber\\
h^a_i &=& -2 g^{ab} k_{i b} + 2 {i \over \sqrt{g}} \e^{ab}(v_{ib}
- \pa_b \chi_i),\nonumber\\
\cg _{ij} &=& k_i^\m k_{j \m},\nonumber \\
\cd _{ij} &=& - (c_{[ij]}+ \l_{[ij]}) - f^k_{\ ij} \chi_k.\label{deffGD}
\eea
After completing the square in the gauge fields $A^i$ the partition
function reads
\bea
Z &=&\int [dA_a^i][dX^i][d\chi]
B[f_i] \det \cf
\nonumber\\
&\qquad& \exp{-\big(S-{1 \over 4 \p \a '} \int d^2 \s \sqrt{g}
h^a_i (f^{-1})^{ij}_{ab} h^b_j\big)}\nonumber\\
&\qquad& \exp{-\big(\frac{1}{4\pi\alpha'}\int d^2\sigma\sqrt{g}f^{ab}_{ij}
A_a^i A_b^j \big)}
\label{dualpartfun5}
\eea

We still have to specify the gauge-fixing functions $f_i$. In contrast
 to the abelian case, specifying a general gauge choice that works
for all gauged sigma-models is still an open problem. In the following
  we will consider, in analogy with the abelian case,
 local gauge choices of the form
\be
\chi_i = \h_i(X^\m),
\ee
where the $\h _i$ are  unspecified  functions on target space. This
means that $f_i= \chi_i  - \h_i$.
Such a choice (locally) fixes the gauge if the Fadeev-Popov
 determinant induced by
it
\be
\det \cf = \det[ \cl_j \h_i + f^k_{\ ij} \h_k ],
\ee
 doesn't vanish.  This  determinant, being independent
of the 2-dimensional metric,  will not contribute to the dilaton
transformation and will be ignored in the following.

A more explicit form for the classical dual action
\be S[\tilde{G}_{\mu\nu}, \tilde{B}_{\mu\nu}, \Phi]=S-{1 \over 4 \p \a '}
\int d^2 \s \sqrt{g}h^a_i (f^{-1})^{ij}_{ab} h^b_j ,
\ee
can be found by
inverting the matrix $f$:
\be
(f^{-1})^{ij}_{ab} = g_{ab} ( \cg - \cd \cg ^{-1} \cd) ^{-1 \ ij}
- {i\epsilon_{ab} \over \sqrt{g}} [\cg^{-1} \cd (\cg - \cd \cg^{-1}
\cd)^{-1}]^{ij}
.\nonumber
\ee
After some algebra we obtain the following form for the dual metric $\tilde{G}
_{\m \n} $ and torsion $\tilde{B} _{\m \n}$:
\bea
\tilde{G} _{\m \n} &=& G_{\m \n} - [k_{i \m} k_{j \n} - (v_{i \m}
  -
\pa_\m \h_i)(v_{j \n} - \pa_\n \h_j)]( \cg - \cd \cg^{-1} \cd) ^{-1 \ ij}
\nonumber \\
&\ &+ 2 k_{i \m}(v_{j \n} - \pa_\n \h_j)[\cg^{-1} \cd (\cg - \cd \cg^{-1}
\cd)^{-1}]^{ij},\nonumber \\
\tilde{B} _{\m \n} &=& B_{\m \n} + [k_{i \m} k_{j \n} - (v_{i \m}
 -
\pa_\m \h_i)(v_{j \n} - \pa_\n \h_j)][\cg^{-1} \cd (\cg - \cd \cg^{-1}
\cd)^{-1}]^{ij}
\nonumber \\
&\ & -2 k_{i \m}(v_{j \n} - \pa_\n \h_j)( \cg - \cd \cg^{-1} \cd) ^{-1 \ ij}.
\eea

\subsection{Regularization}
We still have to incorporate the contribution from the integral over the gauge
fields
\be
\int [dA_a^i]\exp{-\big(\frac{1}{4\pi\alpha'}\int d^2\sigma\sqrt{g}f^{ab}_{ij}
A_a^i A_b^j \big)}.
\label{gfaction}
\ee
A suitable choice of path-integral measure for the gauge
fields will reduce this calculation to the one performed in the
abelian case. The $A^i_a$ can be arranged in a column vector
$$
\vec{A} = \left( \begin{array}{c} A^i_1 \\ A^j_2
\end{array} \right) ;\ i,j=1 \ldots n ,
$$
so that (\ref{gfaction}) now reads
\be
\int [dA_a^i]\exp{-\big(\frac{1}{4\pi\alpha'}\int d^2\sigma\sqrt{g}
\vec{A}^T f \vec{A} \big)}.
\ee
The matrix $f$ should now be seen as a $2n \times 2n$ matrix
$$
f = \left( \begin{array}{cc} \cg & i\cd \\ - i\cd & \cg \end{array} \right) .
$$
The eigenvalues of  matrix $f$ are (at least) twofold
degenerate. Indeed, making
use of the  relation
$$
\det \left( \begin{array}{cc} A & B \\ C & D \end{array} \right)
= \det A D  \det ( 1 - D^{-1} C A^{-1} B) ,
$$
one easily shows that the  characteristic polynomial $\det ( f - \l )$
is a complete square:
\bea
\det ( f - \l) &=& \det(\cg - \l)^2 \det( 1 - ((\cg - \l)^{-1} \cd )^2)\nonumber\\
&=& \det(\cg - \l + \cd ) \det(\cg - \l -\cd )\nonumber\\
&=& \det(\cg - \l + \cd ) \det(\cg - \l + \cd )^T\nonumber\\
&=& (\det(\cg + \cd - \l))^2.
\eea
This means that there exists a matrix $R$ such that
$$
R f R^{-1} = \mbox {diag} (\l_1,\l_2,\cdots ,\l_n,\l_1,\cdots,\l_n)\;, $$
where $\l_i$ are the eigenvalues of the matrix $\cg + \cd $.
Defining
$$
\vec{\tilde{A}} = R \vec{A}
$$
we can, as in (\ref{Aparam}), locally define scalars $\a^i, \b^i$ such that
$$
\tilde{A}^i_a = \partial_a \a^i + {\e_a^{\ b} \over \sqrt{g}\l_i} \partial_b
\b^i
$$
(\ref{gfaction}) now becomes
$$
\int [dA_a^i]\exp{-\big(\frac{1}{4\pi\alpha'}\int d^2\sigma\sqrt{g}g^{ab}
\sum_i (\l_i  \partial_a\alpha^i \partial_b \a^i + {1 \over \l_i}
 \partial_a\b^i \partial_b \b^i)
 \big)}.
$$
The calculation now proceeds precisely as in the abelian case. We introduce
sets of
PV fields $Y_{r0}^i,\  Y_{r1}^i,\ Y_{r2}^i$ with kinetic terms
\begin{eqnarray}
S_{PV}[ Y^i_{r0}]&=&\frac{1}{4\pi\alpha'}\sum_r\int d^2\sigma \sqrt{g}
\, \frac{1}{\l_i}(g^{ab} \partial_a Y^i_{r0} \partial_b Y^i_{r0} + M_{ri}^2
(Y^i_{r0})^2) ,\nonumber\\
S_{PV}[Y^i_{r1}]&=& \frac{1}{4\pi\alpha'}\sum_r\int d^2\sigma \sqrt{g}
\, \l_i(g^{ab} \partial_a Y^i_{r1} \partial_b Y^i_{r1} + M_{ri}^2
(Y^i_{r1})^2) ,\nonumber\\
S_{PV}[Y^i_{r2}]&=&\frac{1}{4\pi\alpha'}\sum_r\int d^2\sigma \sqrt{g}
\, {1 \over\l_i}(g^{ab} \partial_a Y^i_{r2} \partial_b Y^i_{r2} + M_{ri}^2
(Y^i_{r2})^2),
\end{eqnarray}
and the path integral measure is defined as
\begin{eqnarray}
&&[dX^i][dA_a^i][d\chi_i] \nonumber
\\
&=& \prod_\sigma
 \Big( d(\alpha^i (\sigma))d(\beta^i(\sigma)) d(\chi
(\sigma))d(X^i(\s))\nonumber \\
&&\prod_r d(\ Y^i_{r0} (\sigma)) d(Y_{r1}^i(\sigma))d(Y_{r2}^i(\sigma))
\Big).
\end{eqnarray}
Again we neglected the overall Jacobian factor for the change of variables
from $A_a^i$ to $(\a^i,\ \b^i)$ since it would drop out of the final result
anyway.
\subsection{Quantum contributions}
\subsubsection{The original model}
The original model is obtained by integrating over the gauge fields and
the Lagrange multipliers. We choose the gauge $\a^i = 0$. As in
(\ref{W}) we
encounter a contribution $e^{-W}$ with
\be
W =  {1 \over 4 \p} \int d^2 \s \sqrt{g}\sum_i [ {1 \over 12}
R \Box^{-1} R + {1 \over 4}\Box^{-1} R\, \L_i^2 + {1 \over 2} R \ln \l_i
+ 2 c_1 \L_i^2],
\ee
with $\L_{i\ a} = \pa_a \l_i$.
\subsubsection{The dual model}
The same calculation as the one leading to (\ref{tildeW}) gives the dual
 contribution $e^{-\tilde{W}}$ with
\be
\tilde{W} =  {1 \over 4 \p} \int d^2 \s \sqrt{g}\sum_i [ {1 \over 12}
R \Box^{-1} R + {1 \over 4}\Box^{-1} R\,  \L_i^2 + 2 c_1 \L_i^2].
\ee
\subsubsection{Dilaton shift}
Combining contributions from the original and dual side
we arrive at the final result
\begin{eqnarray}
&& \int [dX^i]\prod_{i}[dY^i_1]\,\exp (-S[G_{\mu\nu},B_{\mu\nu},\Phi] -
S_{PV}(Y^i_1))\nonumber\\
&=&\int [dX^i]\prod_{i}[dY^i_0]\;\exp (- S[\tilde
G_{\mu\nu},\tilde B_{\mu\nu},\tilde \Phi]- S_{PV}({Y}^i_0)),
\end{eqnarray}
where the dual dilaton is given by
\bea
\tilde{\F} &=& \F - \half \ln \ \det (\cg + \cd)\nonumber\\
&=& \F - {1 \over 4} \ln \ \det f,
\eea
where $f$, $\cg $ and $\cd $ are defined in (\ref{deffGD}).

This agrees with the dilaton transformation usually adopted in the literature
\cite{quevedo}, \cite{review1}.
The case of commuting isometries can easily be obtained by putting $\cd = 0$.
One then recovers
the abelian results obtained in section 3.

\setcounter{equation}{0}
\section{Conclusions and outlook}

In this paper, we presented an unambiguous derivation of the dilaton
transformation under T-duality with respect to both  abelian and non-abelian
groups of isometries. Although finding the  dual metric and torsion is
easy enough, obtaining the dual dilaton is subtle since care is needed in
defining a correct regularization and measure for the path integral.
An important ingredient in our discussion was the
use of Pauli-Villars regularization. This method consisting of adding
extra
terms to the action  is a very convenient way to fix the regulators
once and for all and to keep track of them in subsequent calculations.

The main observation is that a careful definition of the path integral measure 
for the gauge fields leads to a nontrivial (background-dependent) contribution 
even when going to the original model. This implies that the dilaton 
transformation comes from a ratio of functional determinants, one from the 
original and one from the dual side. From this we also deduce that the dilaton 
transformation itself is a regularization independent effect, since different 
regulators would give extra contributions on both sides that cancel in the aforementioned 
ratio of determinants. 

In section two we presented a gauge fixing that leads to the target-space
covariant form of the Buscher rules for abelian duality. This made it
possible,
for the first time, to derive the dilaton shift
using a manifestly covariant calculation. The key ingredient of our
construction
was a suitable
gauge choice which has no straightforward generalisation to the non-abelian
case.
Though in this case specific
examples have been treated in the literature (see e.g. \cite{quevedo}, 
\cite{alvarez}, \cite{Sfetsos}, \cite{Giveon}),
an explicit general gauge-fixing procedure is not known, making a general
discussion
of the  dual geometry difficult. It might prove useful to study this problem
in a restricted class of backgrounds such as the WZW models or homogeneous
spaces\cite{progress}.

Finally we showed in our discussion of non-abelian duality that gauging is
somewhat less
restricted than is generally assumed, (see
(\ref{gaugecond})) in
the literature. At this point the geometrical meaning of this condition
is not obvious, nor is it clear whether it permits the gauging of
interesting models which have not been considered previously. However it might
very well lead to further insights in the gauge fixing procedure for the
non-abelian case.

\noindent{\Large \bf Aknowledgments}\\
This work was supported in part
by the European Commission TMR programme ERBFMRX-CT96-0045 in
which AS is associated to K.U. Leuven.

\setcounter{equation}{0}
\appendix
\section{Calculation of regularized traces}
\label{app1}

In this appendix we calculate the regularized traces appearing the derivation
of the dilaton transformation. Consider a second order differential
operator of the form\footnote{Epsilon tensors are defined by
 $\epsilon^{ab}\epsilon_{b}^{\;c}=-gg^{ac},\  \varepsilon^{ab}
\varepsilon_{b}^{\;c}=-\d^{ac}$}:

\be
\co (v_i) = \left( \begin{array}{cc}
 \d^{ab}(-  \pa_a \pa_b + v_{1a} \pa_b + v_{2a}v_{2b}) & \varepsilon^{ab}
 v_{3a} \pa_b \\
- \varepsilon^{ab} v_{3a} \pa_b & \d^{ab}(-  \pa_a \pa_b + v_{4a} \pa_b +
v_{5a}v_{5b})
\end{array} \right)
\ee
 where the $v_i \ i = 1 \ldots 5$ are arbitrary world-sheet vectors.
We are interested in a  quantity $W(\r, v_i )$ with the property that
\be
{\d W(\r, v_i ) \over \d \r}= - \sum_j  \frac{c_j}{2}{\rm tr}  \int_0
 ^\infty d\lambda e^{-\lambda}  \sum_n
 \psi_n^* (\sigma) e^{- \frac{\lambda}{M_j^2} \rho^{-1} \co}  \psi_n
(\sigma)
\ee
where ${\rm tr}$ stands for a trace of 2 by 2 matrices, and the $\psi_n$
constitute
 an orthonormal basis for the curved metric: $\int d^2 \sigma \rho \psi^*_n
\psi_m = \delta_{nm}$.
The kernel
\begin{equation}\sum_n \psi_n^* (\sigma) e^{- \frac{\lambda}{M_j^2} \rho^{-1}
\mathcal{O}}
\psi_n (\sigma) \equiv K(\frac{\lambda}{M_j^2}, \sigma, \sigma, \rho^{-1}
\co)\end{equation}
can, following Gilkey \cite{gilkey}, be expanded for large $M_j$ as follows:
\begin{equation}K(\frac{\lambda}{M_j^2}, \sigma, \sigma, \rho^{-1} \co) = E_0
\frac{M_j^2}{\lambda} + E_2 + \ order(M_j^{-2})\,. \end{equation}
In casu,
\bea
tr E_0 &=& \frac{1}{2 \pi}\nonumber\\
tr E_2 &=& {1 \over 4 \p}\left(-v_2^2 - v_5^2 - {1 \over 4} (v_1^2 + v_4^2 -
2v_3^2) + \half g^{ab} \pa_a (v_{1b} + v_{4b}) - {R^{(2)} \over 3}\right)
 \nonumber
\eea
so that, using the relations (\ref{PVcond}) we obtain
\footnote{One may wonder where the usual counterterms,
to be divergent when $M_i^2\rightarrow\infty$, have gone.
Remarkably, some of these have been swept under the carpet by
interchanging the trace and the $\l$ integral.
The proper computation (without this unwarranted interchange) yields an
extra divergent term $\frac{1}{4\pi}\sum_i c_iM^2_i\log  M^2_i\,$.
These terms eventually cancel in the final result and, staying with tradition,
we will ignore them from now on.}

\be
{\d W(\r, v_i ) \over \d \r}=
{1 \over 4 \p}\left({R^{(2)} \over 6} + {1 \over 2}(v_2^2 + v_5^2) + {1 \over
 8}(v_1^2 + v_4^2 - 2v_3^2) - {1 \over 4} g^{ab} \pa_a (v_{1b} +
v_{4b})\right).
\ee

Using the conformal gauge expression for the world-sheet curvature, $R
= - \r^{-1} \d^{ab} \pa_a \pa_b \ln \r$,
this can be integrated to give
\bea
W(\r, v_i) &=& {1 \over 4 \p} \int d^2 \s \sqrt{g} \Big( {1 \over 12}
R \Box^{-1} R + \Box^{-1} R ( + {1 \over 2}(v_2^2 + v_5^2)\nonumber\\
&\ & + {1 \over
 8}(v_1^2 + v_4^2 - 2v_3^2) - {1 \over 4} g^{ab} \pa_a (v_{1b} + v_{4b}))
\Big) + {\rm Weyl \ inv. terms}
\eea
We can now apply this general formula to the following special cases:
\begin{itemize}
\item In going to the original model, one encounters the operator $\co$
corresponding to $v_1 = -\L; \ v_4 = - \L; \ v_2 = v_3 = v_5 = 0$.
This gives a contribution
\be
W =  {1 \over 4 \p} \int d^2 \s \sqrt{g} \left( {1 \over 12}
R \Box^{-1} R + {1 \over 4}\Box^{-1} R \L^2 + \half R \ln G_{00}\right)
 + {\rm Weyl \ inv. terms}
\label{origcontrib}
\ee
\item On the dual side, the operator $\tilde{\co}$ corresponds to
$v_1 = - \L, \ v_4 = \L; \ v_2 = v_3 = v_5 = 0$, leading  to
\be
\tilde{W} =  {1 \over 4 \p} \int d^2 \s \sqrt{g} \left( {1 \over 12}
R \Box^{-1} R + {1 \over 4}\Box^{-1} R \L^2 \right) + {\rm Weyl \ inv. terms}.
\label{dualcontrib}
\ee
\item The calculation indicated by Buscher in \cite{Buscher} makes use of a regulator
with $v_1 = v_4 = v_3 = - \L; \ v_2 = v_5 = {\L \over 2}$. The full 
contribution (after partial integration) reads:
\be
W(\r, v_i)  = {1 \over 4 \p} \int d^2 \s \sqrt{g} \left( {1 \over 12}
R \Box^{-1} R + {1 \over 4}\Box^{-1} R \L^2 - {1 \over 2} R \ln G_{00}\right)
 + {\rm Weyl \ inv. terms}.
\label{Buscherresult}
\ee
The term proportional to $\L^2$ is missing in \cite{Buscher}.
\end{itemize}

\end{document}